\documentclass[prl,twocolumn,10pt,aps]{revtex4-1}

\usepackage[utf8]{inputenc}
\usepackage{amssymb}
\usepackage{amsmath}
\usepackage{graphicx}
\usepackage{subfigure}
\usepackage{dcolumn}
\usepackage{bm}
\usepackage{hyperref}
\usepackage{enumerate}
\usepackage{xcolor}
\usepackage{natbib}
\usepackage{amssymb}
\usepackage{amsbsy}
\usepackage{graphics}
\usepackage{comment}
\usepackage{hyperref}

\begin{document}

\preprint{APS/123-QED}

\title{Scalar signatures of chaotic mixing in porous media}

\author{J. Heyman}
\affiliation{Geosciences Rennes, UMR 6118, Universit\'{e} de Rennes 1, CNRS, 35042 Rennes, France}
\author{D. R. Lester}
\affiliation{School of Engineering, RMIT University, 3000 Melbourne, Australia}
\author{T. Le Borgne}
\affiliation{Geosciences Rennes, UMR 6118, Universit\'{e} de Rennes 1, CNRS, 35042 Rennes, France}
\email{joris.heyman@univ-rennes1.fr}

\date{\today}

\begin{abstract}
Steady laminar flows through porous media spontaneously generate Lagrangian chaos at pore scale, with qualitative implications for a range of transport, reactive and biological processes. The characterization and understanding of mixing dynamics in these opaque environments is an outstanding challenge. We address this issue by developing a novel technique based upon high-resolution imaging of the scalar signature produced by push-pull flows through porous media samples. Owing to the rapid decorrelation of particle trajectories in chaotic flows, the scalar image measured outside the porous material is representative of \textit{in-situ} mixing dynamics. We present a theoretical framework for estimation of the Lypapunov exponent based on extension of Lagrangian stretching theories to correlated aggregation.  This method provides a full characterization of chaotic mixing dynamics in a large class of porous materials.
\end{abstract}


\maketitle


Recent experimental~\cite{Kree2017,Heyman2020,Souzy:2020aa}, numerical~\cite{Turuban2018} and theoretical~\cite{Lester2013} results have shown that the topological complexity inherent to three-dimensional (3D) porous media generates chaotic advection at the pore scale. This means that, in steady laminar flows, fluid elements are elongated at an exponential rate, qualitatively impacting~\cite{aref2017frontiers} the transport, mixing and reactivity of solutes, colloids and particles. Although well documented in dynamical systems~\cite{TelPhysRep2005}, the consequences of chaotic mixing are yet to be uncovered in porous substrates. In particular, chaotic advection enhances chemical gradients at microscale over a large range of P\'eclet numbers~\cite{Heyman2020}, a phenomenon that may explain the limitations of conventional macrodispersion models to predict reactive transport~\cite{deAnnaEST2014}. 
\begin{figure}[h!]
\centering \includegraphics[width=\linewidth]{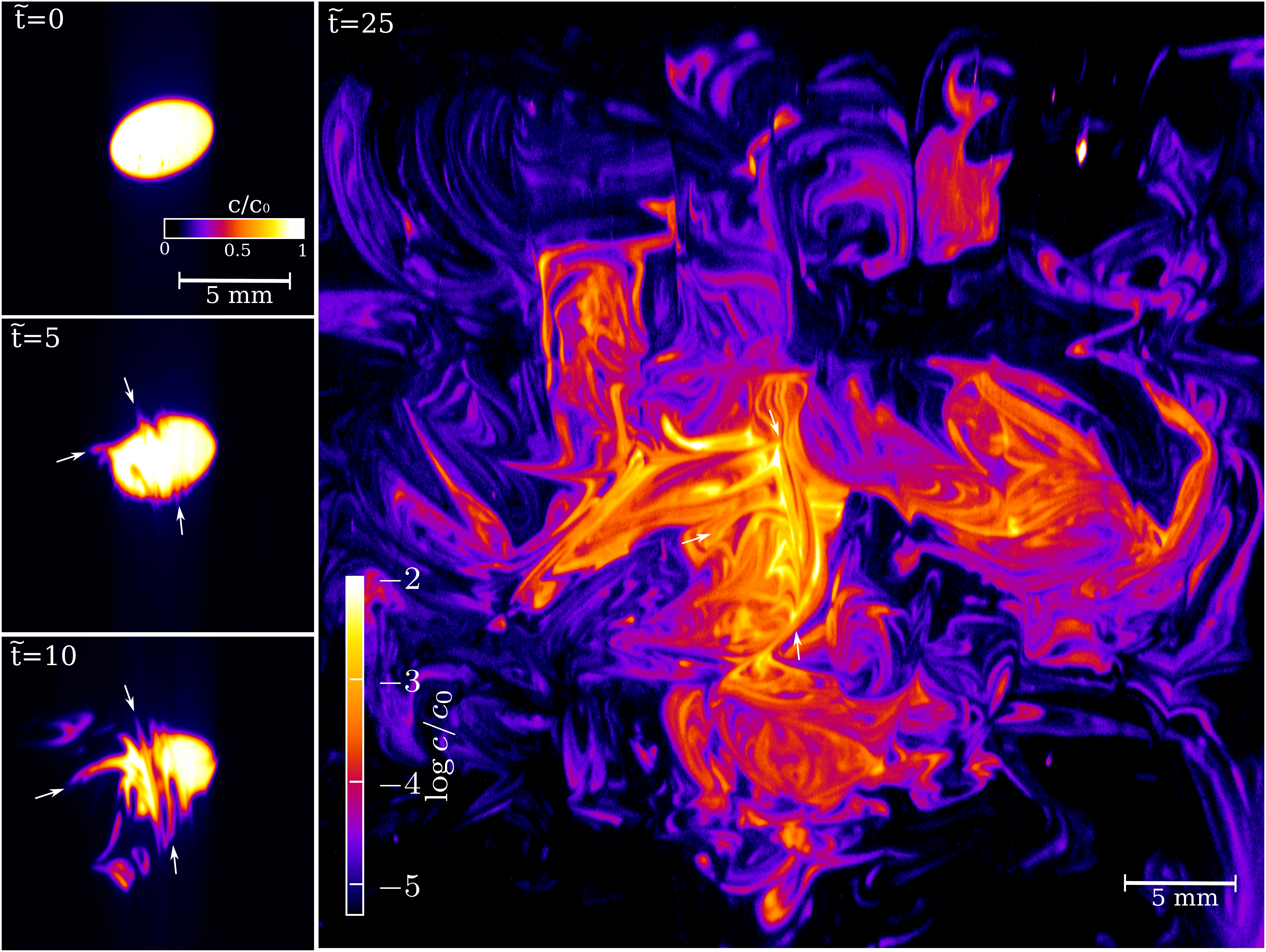}
\caption{Heterogeneous and anisotropic scalar signature of the push-pull echo during a push-pull experiment through random bead packings. The initially spheroidal dye blob at $\tilde{t}=t^\prime\bar{u}/d=0$ diffuses preferentially along directions of high fluid deformation (indicated by arrows), leading to a highly striated dye distributions at later times. See video in \cite{supp-mat}.}\label{fig:leakage}
\end{figure}
Dispersion models are based on the description of the spatio-temporal spread of transported solutes~\cite{bijeljic2011signature}, but do not capture the scalar heterogeneity inside the plume. These mixing processes evolve with distinct dynamics which remain poorly understood in porous media~\cite{dentz2011mixing,LeBorgneJFM2015}. Experimental characterisation of mixing dynamics in porous media was achieved in two-dimensional (2D) micromodels that facilitated the imaging of scalar gradients below pore-scale~\cite{deAnnaEST2014}. However, 2D pore topologies prohibit the development of chaotic trajectories in steady flows~\cite{Lester2013}. In turn, 3D imaging of solute transport was achieved with X-ray and magnetic resonance techniques~\cite{hunter2007nmr,boon2016}, although the resolution is generally unable to resolve the whole spectrum of microscale scalar gradients~\cite{aref2017frontiers,Heyman2020}. While offering better resolution, refractive index matching techniques~\cite{Kree2017,Souzy:2020aa,Heyman2020} are limited to optically transparent materials. Thus, quantification of chaotic mixing dynamics across the diversity of natural and industrial porous matter remains an outstanding challenge.

Here we develop a novel methodology based on \textit{push-pull} experiments to provide quantitative measurements of chaotic mixing in opaque porous media. Also termed \textit{echo} experiments~\cite{hulin1989echo}, conventional push-pull experiments consist of two phases. First, a pulse of solute dye is injected into a porous sample via a steady Stokes flow (push phase). Then, the flow is reversed (pull phase) and the withdrawn solute mass is monitored at the injection point. This scalar echo provides an indirect quantification of solute longitudinal dispersion~\cite{hulin1989echo}, but does not capture mixing processes~\cite{dentz2011mixing}. To characterize the entire mixing history, we use high-resolution imaging of the withdrawn solute spatial signature in the plane transverse to the mean flow direction (Fig.~\ref{fig:leakage} and \cite{supp-mat}).
Despite the reversibility of Stokes flows, chaotic advection coupled to molecular diffusion induces fast decorrelation of solute paths, a general property of chaotic flows~\cite{Slipantschuk:2013aa}. This decorrelation renders the asymptotic scalar echo statistically equivalent to that of two consecutive push-only flows and enables the indirect quantification of \textit{in-situ} mixing dynamics in porous media. In the limit of low dye molecular diffusivity $D_\text{m}$, we show that the average fluid stretching rate can be obtained from the temporal decay of the spatial variance $\sigma^2_c(t)$ of the scalar echo. These results are supported by numerical simulations~\cite{supp-mat} of scalar mixing in the Sine Flow, a prototype of chaotic advection in fluids~\cite{Meunier2010,haynes2005controls}, and confirms the universality of our findings.

\begin{figure}[h]
\centering \includegraphics[width=\linewidth]{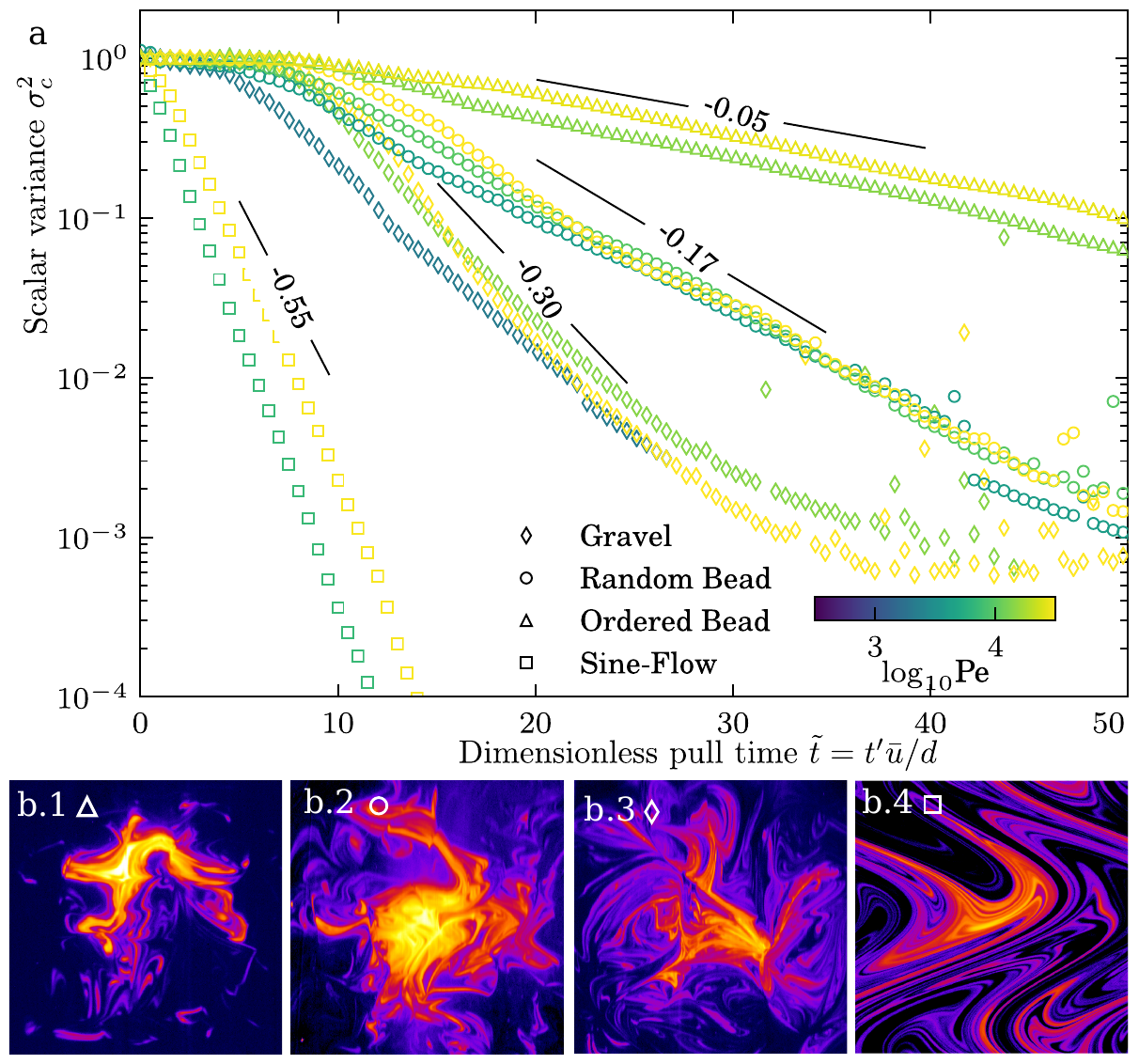}
\caption{a. Spatial variance of scalar echo $\sigma_c^2$ as a function of time since reversal (expressed as pore advection time $t^\prime \bar{u} / d$) for various P\'{e}clet numbers and the three class of porous media considered. The noise level of the camera produces a saturation of scalar variance at a level $\langle c^2 \rangle \approx 10^{-3}$. b. Snapshots of scalar concentration in the pull phase (in logarithmic scale) at $25 \, \tilde{t} \bar{u}/d$ for (b.1) ordered bead pack,  (b.2) random bead pack, (b.3) gravel pack and (b.4) Sine Flow (at $t^\prime=10$). Videos are available in \cite{supp-mat}.}\label{fig:variance}
\end{figure}

Push-pull experiments are carried out in granular columns containing random and ordered (body-centered cubic) packings of monodispersed beads of diameter $d=5$~mm or gravels of mean grain diameter $d=5.4$~mm. A steady Stokes flow with mean longitudinal velocity $\bar{u}$ of a viscous glycerol-water mixture is created via a constant pressure gradient between the column extremities (Fig. \ref{fig:variance}b). The experiment starts by continuously injecting a fluorescent solute dye (molecular diffusivity $D_\text{m}\approx 10^{-11}$m$^2$s$^{-1}$ in the glycerol-water mixture) through a small needle ($r=0.5$~mm) at the top of the packing (push phase), until a steady solute plume has formed inside the porous media. We then stop the injection and smoothly reverse the flow by inverting the pressure gradient (pull phase), while a camera records the spatial distribution of dye concentration at the injection plane, via a thin laser sheet sectioning the flow transversally (see \cite{supp-mat} for details on the setup). The continuous injection of the dye (versus pulsed in classical push-pull experiments) strongly reduces longitudinal scalar gradients driving longitudinal dispersion and allows focusing on transverse chaotic mixing dynamics.

As Stokes flows are linear and time-reversible, fluid elements experience zero net deformation over the complete push-pull cycle. Thus, the distribution of dye molecules shown in Fig.~\ref{fig:leakage} at time $t^\prime$ after flow reversal have travelled an average distance $\ell \sim \bar{u}t^\prime$ from the injection point into the porous sample before being withdrawn to the injection plane. In a simple translational flow, these molecules would return to the injection within a small diffusive radius $r_D\sim\sqrt{4D_\text{m}t^\prime}$ of their initial position. The circular scalar footprint of the plume formed at the injection point (Fig.~\ref{fig:leakage}, $\tilde{t}=0$) would thus be globally maintained upon flow reversal. Conversely, in chaotic flows, fluid deformation over the push-pull cycle significantly amplifies the effective mixing of the withdrawn solute, as was observed~\cite{oxaal1994irreversible} for the push-pull flow over a stagnation point. Sustained exponential stretching of fluid material elements caused by chaotic advection renders dispersion strongly heterogeneous and anisotropic, as evidenced by the highly striated scalar distribution imaged experimentally in the pull phase (Fig.~\ref{fig:leakage}), which are generated by directions of exponential amplification (retardation) of diffusion along the unstable (stable) manifolds of the chaotic flow. Owing to flow stationarity, these manifolds are time-invariant and so the scalar echo converges towards a steady spatial structure (see Video in \cite{supp-mat}).
After a short transient phase, the spatial variance $\sigma^2_c$ of this structure exhibits an exponential decay $\sigma^2_c \sim e^{-\gamma_2 \tilde{t}}$ (Fig~\ref{fig:variance}a), where $\tilde{t}= t^\prime \bar{u}/d$ is the dimensionless pull time and $d$ the mean grain diameter. At late times, the scalar variance tends to a constant $\sigma^2_c \approx 10^{-3}$ corresponding to the background noise level of the camera. This exponential decay is also consistently observed in numerical simulations~\cite{supp-mat}, where the noise level is much lower. The decay exponent $\gamma_2$ is independent of the Péclet number ($\text{Pe}=\bar{u}d/D_\text{m}$) over the range $10^3$ to $10^4$, but strongly varies with the porous medium properties, with significantly higher values for gravels ($\gamma_2=0.30$) than for random bead packings ($\gamma_2=0.17$) and ordered packings ($\gamma_2=0.05$).

To relate the exponent $\gamma_2$ to the characteristics of chaotic mixing, we adopt the Lagrangian Stretching (LS) framework~\cite{balkovsky1999universal,Villermaux2019}. This theory considers the balance of flow stretching and molecular diffusion transverse to the elongated lamellar scalar structures formed during fluid deformation in push-only flows (see Fig. 2 in \cite{supp-mat}). For a constant stretching rate $\lambda$, an isolated lamella of length $\l$ elongates as $\rho(t)=\l(t)/l(0)=e^{\lambda t}$ while its width decay as $\rho^{-1}$ due to fluid incompressibility. Once the width reaches the Batchelor scale $s_B=\sqrt{D_\text{m}/\lambda}$~\cite{supp-mat}, stretching and diffusion balance each other out and the lammela peak concentration $c_\text{max}$ asymptotically decays as $\rho^{-1}$~\cite{supp-mat}. The average squared concentration over an area $\mathcal{A}$ then follows~\cite{Meunier2010,supp-mat}
\begin{equation}
 \overline{c^2} \propto \mathcal{A}^{-1} \l s_B c_\text{max}^2 \sim \rho^{-1},
\label{eq:isolated}
\end{equation}
where the overbar denotes two-dimensional spatial average. Owing to solute mass conservation, $\overline{c}$ is constant and the evolution of $\sigma^2_c$ is driven by $\overline{c^2}$. As shown in~\cite{supp-mat}, this scaling accurately captures the scalar echo produced by a single stagnation point~\cite{oxaal1994irreversible}. This is because enhancement of diffusion by fluid stretching only occurs during the push phase. In the pull phase, scalar gradients are unaltered as they remain perpendicular to the main compression direction.
In contrast, folding of material lines in chaotic flows induces ergodicity and a systematic decorrelation~\cite{supp-mat} of solute particle trajectories and stretching directions between the push and pull phases over a timescale of order of $\lambda^{-1}$~\cite{Slipantschuk:2013aa}.   
Thus, for $t\gg \lambda^{-1}$, the stretching histories of fluid elements during the push and pull phases are independent and the scalar echo observed at time $t^\prime$ is equivalent to the mixing produced by a push-only flow of duration $t = 2 t^\prime$, as verified from numerical simulations of push-pull flows in the Sine Flow~\cite{supp-mat}. 

The evolution of the scalar variance of an isolated lamella stretched at a constant rate (Eq.~\ref{eq:isolated}) is generalizable to randomly varying stretching rates, as typically experienced by fluid elements at pore-scale~\cite{Lester2013}. Because of the multiplicative nature of stretching, the log-elongation of material elements $\log \rho$ is well approximated in ergodic chaotic flows by a sum of iid random variables, that converges towards the normal distribution with mean $\bar{\lambda} t$ and variance $\sigma^2_{\lambda} t$~\cite{Meunier2010}. Ensemble averaging (denoted by angled brackets) of $\eqref{eq:isolated}$ over this distribution reads
\begin{equation}
\sigma^2_c \sim \left \langle \overline{c^2}  \right \rangle \sim \int_{0}^{\infty}  \text{d}\Lambda \, \exp\left(- t  \frac{(\Lambda-\bar{\lambda})^2}{2\sigma^2_{\lambda} } - \Lambda t\right),
\label{eq:int1}
\end{equation}
with $\Lambda=\log\rho/t$.
At large times, this integral can be approximated with the Laplace method. The term dominating the integral is $\exp\left(-t(\Lambda^*-\bar{\lambda})^2/(2\sigma^2_{\lambda} ) - \Lambda^*t\right)$, where the saddle point is $\Lambda^* \equiv \text{max}\left(\bar{\lambda}-\sigma^2_{\lambda}  , 0 \right)$ \cite{balkovsky1999universal,haynes2005controls}. We verified this approximation numerically.
For smooth and space-filling flows, $\bar{\lambda}\approx \sigma_\lambda^2$~\cite{Meunier2010}, hence $\Lambda^*=0$ and $\sigma^2_c \sim e^{-\bar{\lambda}^2/(2\sigma^2_{\lambda} ) t} \approx e^{-(\bar{\lambda}/2) t}$, e.g. the decay exponent is determined by the fraction of lamellae that have experienced no stretching and is independent of the P\'eclet number.

The validity of the LS theory for isolated lamellae extends well beyond coalescence time, for flows in the Batchelor regime~\cite{haynes2005controls}, e.g. when the scalar fluctuations lengthscales $l_c$ are smaller than the velocity lengthscales $l_v$. 
Such regime naturally develops in porous media at high P\'eclet number, for which $s_B \ll l_v$~\cite{Heyman2020}. This unexpected persistence was mathematically associated~\cite{haynes2005controls} to the existence of a continuous limit in the spectrum of the scalar covariance equation, when Pe$\to\infty$. We found that it can also be explained by the dominance of \textit{correlated} aggregation of lamellae in the Batchelor regime. In this regime, coalescing lamellae form bundles that remain smaller than the velocity fluctuation lengthscale ($l_c \leq l_v$). Thus, fluid stretching is approximately uniform within a bundle. In turn, the number $N$ of lamellae in the bundle is dictated by local fluid compression and therefore proportional to the local fluid elongation $\rho$ in incompressible flows.
Thus, weakly stretched and compressed fluid elements are also weakly aggregated  ($N \sim \rho$~\cite{supp-mat}), a correlated aggregation scenario verified numerically in the Sine Flow~\cite{supp-mat}. As the fraction of weakly stretched lamellae  ($\rho \approx 1$) asymptotically dominates the integral $\eqref{eq:int1}$, and since this fraction is also weakly aggregated  ($N \approx 1$), the domain of validity of Eqs.~\eqref{eq:c2decay}--\eqref{eq:int1} extends beyond coalescence time. 

Recalling that the stretching histories between the push and the pull flows are independent at late time ($t\approx 2t^\prime$), the asymptotic scalar echo variance then follows
\begin{equation}
\sigma_c^2 \sim  e^{- \gamma_2  t^\prime}, \text{ with } \gamma_2 \approx \tilde{\lambda} = \bar{\lambda}d/\bar{u}.
\label{eq:c2decay}
\end{equation}
$\tilde{\lambda}$ is the infinite-time Lyapunov exponent of the porous flow made dimensionless by the pore advection time $t_a=d/\bar{u}$. Eq.~\ref{eq:c2decay} exhibits excellent agreement (Fig. \ref{fig:variance}a) with the Sine Flow simulations for $\text{Pe}=10^3$ and $10^4$, where the Lyapunov exponent $\bar{\lambda} \approx 0.55$ was computed independently~\cite{supp-mat}. The measure of scalar dissipation in push-only flows~\cite{Heyman2020} through index-matched random bead packs at $\text{Pe}\approx 10^4$ yields $\tilde{\lambda}=0.18$~\cite{supp-mat}, in excellent agreement with the push-pull estimate $\tilde{\lambda} \approx 0.17$ from this study. This value is also close to the stretching and folding model proposed~\cite{Heyman2020} for granular porous media, that predicts $\tilde{\lambda}\approx 0.21$ in random bead packs. Note that \cite{Souzy:2020aa} found larger Lyapunov exponents  based on experimental velocity fields ($\tilde{\lambda}\approx 0.5$), but this estimate integrates both longitudinal and transverse stretching.

The weak dependence of $\gamma_2$ on Péclet number, observed both experimentally and numerically  (Fig. \ref{fig:variance}a), confirms that mixing occurs in the Batchelor regime where aggregation is solely determined by the local stretching of fluid elements. Indeed, the initial tracer injection radius $r$ was chosen such that $ l_c \sim r \ll l_v\sim d$ and the P\'eclet was chosen to be large enough for $s_B$ to be much smaller than $l_v$. In turn, the periodic boundary conditions of the Sine Flow ensures $l_ c \leq l_v \sim1$. In contrast, once the mixed scalar forms patches of uniform concentration over larger scales $l_c \gg l_v$, the aggregation of these patches becomes independent of the local stretching statistics and occurs at random~\cite{DuplatVillermauxPRL2003} at a rate given by large scale dispersive motions. In such limit, the decay of $\sigma^2_c$ has been found~\cite{haynes2005controls} to be strongly dependent on Péclet number. 

Differences in the Lyapunov exponent between gravel and bead packings may be explained by the role of granular contacts in controlling the stretching and folding of material lines~\cite{Heyman2020}.  Given their irregular shapes, gravel packs likely possess a larger number density of contacts than random bead packs, favoring chaotic mixing. Conversely, the small Lyapunov exponent associated with flow through ordered packings may be attributed to the existence of flow barriers imposed by the packing symmetries, which may retard chaotic advection~\cite{Turuban2018}.
\begin{figure}
\centering \includegraphics[width=0.9\linewidth]{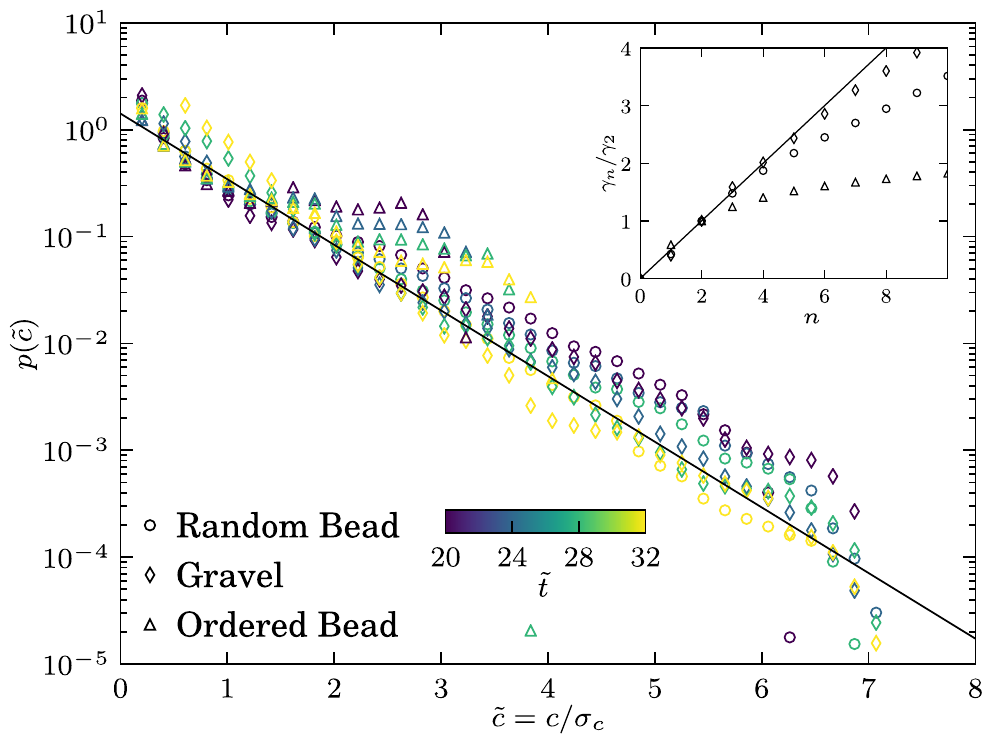}
\caption{Probability density function $p(\tilde{c})$ of scalar concentration rescaled by standard deviation $\tilde{c}\equiv c/ \sigma_c$, for increasing pull time $\tilde{t} = t^\prime \bar{u} /d$ (colors). Points are ensemble averages over several experiments. Straight line indicates the exponential distribution $e^{-\tilde{c}}$. Inset : Exponent $\gamma_n$ for $\left\langle c^n \right\rangle \sim e^{-\gamma_n t} $. Black line stands for the pure self-similar case with $\gamma_n=n\gamma_2/2$.}\label{fig:moments}
\end{figure}
In gravel and random bead packs, the asymptotic exponential decay of the scalar variance is associated with a transition to a statistically stationary scalar probability density function (pdf) with a corresponding exponential distribution $p(\tilde{c})=e^{-\tilde{c}}$, with the rescaled scalar concentration $\tilde{c}\equiv c/\sigma_c$ (Fig.~\ref{fig:moments}). This asymptotic stationary scalar pdf is a common feature of many chaotic flows, such as turbulent flows and random velocity fields~\cite{aref2017frontiers}, and is associated with the emergence of a dominant non-trivial strange eigenmode~\cite{Liu:2004aa} of the advection-diffusion operator. 
Self-similarity is also suggested by linearity of the scalar moments $\langle c^n\rangle$ decay exponents, i.e. $\gamma_n\propto n$, as shown in the inset of Fig.~\ref{fig:moments}. Conversely, the experiments with ordered packings do not exhibit such behavior, probably because the slower mixing rate $\gamma_2$ delays emergence of the dominant eigenmode.

In conclusion, the spatio-temporal imaging of push-pull flows allows the quantification of solute mixing in opaque porous matter, which is currently inaccessible by other techniques. This opens new opportunities to uncover these dynamics in the variety of porous materials that span geologic, biological and engineering applications, where Stokes flows are expected to be chaotic~\cite{Heyman2020}. We established general properties of these flows, including the decorrelation of solute trajectories, the control of scalar dissipation by the correlated aggregation in the Batchelor regime, and the self-similarity of the scalar pdf associated to the dominance of a strange eigenmode. When performed at high P\'eclet numbers, the method allows estimating the Lyapunov exponent, which is the key paramater for chaotic mixing models~\cite{Villermaux2019}. These findings are generic to chaotic flows and are thus relevant to a broad range of fluid applications. 

\begin{acknowledgments}
JH acknowledges financial support by project SUCHY grant ANR-19-CE01-0013, TLB acknowledges financial support by project ReactiveFronts ERC grant 648377.
\end{acknowledgments}

%

\end{document}